\documentclass[aps,twocolumn,showpacs,pra,twoside,amssymb,amsmath]{revtex4}
\usepackage{amssymb}
\usepackage{graphicx}
\usepackage{amsmath}
\usepackage{colordvi}
\usepackage{bbm}
\usepackage{amsmath}

\newcommand{\eins}{\ensuremath{\mathbbm 1}}

\begin{document}

\title{Detection of bound entanglement in continuous variable systems}
\author{Cheng-Jie Zhang$^{1,3}$}
\email{zhangcj@mail.ustc.edu.cn}
\author{Hyunchul Nha$^{2}$}
\author{Yong-Sheng Zhang$^{1}$}
\email{yshzhang@ustc.edu.cn}
\author{Guang-Can Guo$^{1}$}
\affiliation{$^1$Key Laboratory of Quantum Information, University
of Science and Technology of China, Hefei, Anhui 230026,
People's Republic of China\\
$^2$Department of Physics, Texas A \& M University at Qatar, Doha,
Qatar\\
$^3$Department of Physics, National University of Singapore, 2 Science Drive 3, Singapore 117542}

\begin{abstract}
We present several entanglement conditions in order to detect bound entangled states in continuous variable systems. Specifically, Werner and Wolf [Phys. Rev. Lett. 86, 3658 (2001)] and Horodecki and Lewenstein [Phys. Rev. Lett. 85, 2657 (2000)] have proposed examples of bound entangled Gaussian state and bound entangled non-Gaussian state, respectively, of which entanglement can be detected by using our entanglement conditions.
\end{abstract}


\maketitle

\section{Introduction}
Entanglement was recognized as a spooky feature of quantum mechanics in the early days of the 20th century \cite{c1EPR,c1schrodinger,c1Bell}. In recent years, with the progress of quantum information theory, it has become clearer that entanglement is an essential resource in quantum computation and communication \cite{nielsen}. In a related context, the detection of entanglement, for both discrete variables and continuous variables (CVs), is a problem of fundamental importance in quantum information science  \cite{review1,review2,review3}.

There are many efficient criteria proposed for entanglement detection in both discrete variable systems and continuous variable systems \cite{detection1,detection2,detection3,detection4,detection5,Fei,Osterloh,Augusiak,Ou,Baghbanzadeh,Yu,hierarchy,spin,Toth,Peres,LUR1,LUR2,CM,CM2,CM3,nEW,CCN1,CCN,Vogel1,Vogel2,
Guhne,Simon,Duan,Wolf,Giedke,Shchukin,miran,Nha,Giovannetti,multi1,c4gs,c4Nha3,c4Nha4,c4Nha5,c4Hillery,c4serafini,c4son,c4cfrd,c4salles,c4adam}. In discrete variable systems, the well-known positive partial transposition (PPT) criterion is necessary and sufficient for certain low dimensional systems, but it is known to be only necessary for higher dimensions \cite{Peres}. As complementary to the PPT criterion, other criteria have been presented, such as local uncertainty relations \cite{LUR1,LUR2}, covariance matrix criterion (CMC) \cite{CM,CM2}, the computable cross-norm or realignment criterion \cite{CCN1,CCN} and so on. These criteria can be used to detect bound entangled states for which the PPT criterion fails. Recently, Sperling and Vogel proposed a general form of entanglement witness \cite{Vogel1} and negative quasi-probabilities \cite{Vogel2} for arbitrary bipartite entanglement.
In CV systems, Simon and Duan \textit{et al.} presented necessary and sufficient conditions for two-mode Gaussian states in Refs. \cite{Simon} and \cite{Duan}, respectively. Refs. \cite{Wolf,Giedke} improved those results and presented necessary and sufficient conditions for bipartite Gaussian states of arbitrary number of modes. Shchukin and Vogel provided an infinite series of inequalities that is equivalent to the PPT criterion \cite{Shchukin}. Furthermore, Nha and Zubairy proposed separability conditions via uncertainty principle for all negative partial-transpose states \cite{Nha}.

Although numerous entanglement criteria have been proposed for CVs, many of them are corollaries of the PPT criterion, or equivalent to the PPT criterion. For instance, the conditions in Refs. \cite{Simon,Duan,c4gs,c4Nha3,c4Nha4,c4Hillery} are corollaries of the PPT criterion, and the infinite series of inequalities in Refs. \cite{Shchukin,Nha} are equivalent to the PPT criterion. There are not many criteria which can be used to detect bound entangled states for CV systems (some criteria for CV bound entanglement have been proposed, for example, in Ref. \cite{eisert}, CV bound entanglement has been found and experimentally detected and characterized), and one therefore needs other entanglement conditions for CVs to complement the PPT criterion. In this work, we present entanglement criteria in order to detect bound entangled states for CVs.
We show that bound entangled Gaussian state and bound entangled non-Gaussian state proposed in Refs. \cite{Wolf,Horodecki2000} can be detected using our criteria.

The paper is organized as follows. In Sec. II we propose several entanglement conditions for CVs and detect bound entangled Gaussian and non-Gaussian states using our conditions. In Sec. III we discuss the extension of our conditions together with open questions, and give a brief summary of our results.

\section{Entanglement conditions for continuous variables}
Before embarking on our conditions, let us briefly review the criteria in Refs. \cite{Duan,Giovannetti}. Duan \textit{et al.} proposed an inequality for separable states \cite{Duan},
\begin{equation}\label{Duan}
    \langle(\Delta\hat{u})^{2}\rangle+\langle(\Delta\hat{v})^{2}\rangle\geq
a^{2}+1/a^{2},
\end{equation}
where $\hat{u}=|a|\hat{x}_{1}+\hat{x}_{2}/a$ and
$\hat{v}=|a|\hat{p}_{1}-\hat{p}_{2}/a$ with $\hat{x}_{j}$ and $\hat{p}_{j'}$ satisfying $[\hat{x}_{j},\hat{p}_{j'}]=i\delta_{jj'}$ ($j,j'=1,2$).
Moreover, Mancini \textit{et al.} improved the condition and provided a stronger inequality \cite{Giovannetti},
\begin{equation}\label{Mancini}
    \langle(\Delta\hat{u})^{2}\rangle\langle(\Delta\hat{v})^{2}\rangle\geq\frac{1}{4}(|a_1b_1\langle \hat{C}_1\rangle|+|a_2b_2\langle \hat{C}_2\rangle|)^2,
\end{equation}
where $\hat{u}=a_1\hat{x}_1+a_2\hat{x}_2$, $\hat{v}=b_1\hat{p}_1+b_2\hat{p}_2$, $\hat{C}_j=i[\hat{x}_j,\hat{p}_j]$, $a_j$ and $b_j$ are real parameters.
Although the inequalities (\ref{Duan}) and (\ref{Mancini}) are necessary and sufficient conditions for two-mode Gaussian states, it has been proved that both of them are corollaries of the PPT criterion \cite{Shchukin,Nha}. Ref. \cite{TLUR} also proposed a stronger form of the inequality (\ref{Duan}),
\begin{eqnarray}
\langle(\Delta\hat{u})^{2}\rangle+\langle(\Delta\hat{v})^{2}\rangle\geq
a^{2}+\frac{1}{a^{2}}+M^{2}\label{TLUR},
\end{eqnarray}
where $\hat{u}=|a|\hat{x}_{1}+\hat{x}_{2}/a$, $\hat{v}=|a|\hat{p}_{1}-\hat{p}_{2}/a$, and $M=|a|\sqrt{\langle(\Delta\hat{x}_{1})^{2}\rangle+\langle(\Delta\hat{p}_{1})^{2}\rangle-1}-\sqrt{\langle(\Delta\hat{x}_{2})^{2}\rangle+\langle(\Delta\hat{p}_{2})^{2}\rangle-1}/|a|$. In the following, we will present a stronger condition than the inequalities (\ref{Duan}) and (\ref{TLUR}), which can be used to detect bound entangled Gaussian state. Thus, unlike Eqs. (\ref{Duan}) and (\ref{Mancini}), our condition is not a corollary of the PPT criterion.

We first prove two general inequalities, from which our practical conditions can be derived. Choosing arbitrary local operators $\{A_i\}_{i=1}^{n}$ and $\{B_j\}_{j=1}^{m}$ for subsystems A and B, respectively, one can define matrices $R$ and $C$ as $R_{ij}=\langle A_{i}\otimes B_{j}\rangle$, $C_{ij}=\langle A_{i}\otimes B_{j}\rangle-\langle A_i\otimes I\rangle\langle I\otimes B_j\rangle$. The following two inequalities must hold for separable states $\rho=\sum_{k}p_{k}\rho_{k}^{A}\otimes\rho_{k}^{B}$,
\begin{eqnarray}
\|R\|\leq\sum_{k}p_k\sqrt{(\sum_i\langle A_i\rangle_{k}^{2})(\sum_j\langle B_j\rangle_{k}^{2})},\label{R}\\
\|C\|\leq\sqrt{[\sum_i\langle(\Delta A_i)^2\rangle_{\rho_A}-U_A][\sum_j\langle(\Delta B_j)^2\rangle_{\rho_B}-U_B]},\label{C}
\end{eqnarray}
where $\|\cdot\|$ stands for the trace norm (i.e. the sum of the singular values) and the index $k$ refers to the component state $\rho_{k}^{A}\otimes\rho_{k}^{B}$ as $\langle A_i\rangle_k=\mathrm{Tr}\{A_i\rho_{k}^{A}\}$, $\langle B_j\rangle_k=\mathrm{Tr}\{B_j\rho_{k}^{B}\}$. $U_A$ and $U_B$ are non-negative bounds for arbitrary local states as $\sum_i\langle(\Delta A_i)^2\rangle\geq U_A$, $\sum_j\langle(\Delta B_j)^2\rangle\geq U_B$. Inequalities (\ref{R}) and (\ref{C}) can be proved as
$\|R\|=\|\sum_k p_k\mathbf{a}\mathbf{b}^T\|\leq\sum_k p_k\|\mathbf{a}\mathbf{b}^T\|=\sum_{k}p_k\sqrt{(\sum_i\langle A_i\rangle_{k}^{2})(\sum_j\langle B_j\rangle_{k}^{2})}$, and
\begin{eqnarray}
\|C\|&=&\frac{1}{2}\|\sum_{kk'}p_{k}p_{k'}\mathbf{c}\mathbf{d}^{T}\|\nonumber\\
&\leq&\frac{1}{2}\sum_{kk'}p_{k}p_{k'}\|\mathbf{c}\mathbf{d}^{T}\|\nonumber\\
&=&\frac{1}{2}\sum_{kk'}p_{k}p_{k'}\sqrt{\mathrm{Tr}(\mathbf{c}\mathbf{c}^T)\mathrm{Tr}(\mathbf{d}\mathbf{d}^T)}\nonumber\\
&\leq&\frac{1}{2}\sqrt{[\sum_{kk'}p_{k}p_{k'}\mathrm{Tr}(\mathbf{c}\mathbf{c}^T)] [\sum_{kk'}p_{k}p_{k'}\mathrm{Tr}(\mathbf{d}\mathbf{d}^T)]}\nonumber\\
&\leq&\sqrt{[\sum_i\langle(\Delta A_i)^2\rangle_{\rho_A}-U_A][\sum_j\langle(\Delta B_j)^2\rangle_{\rho_B}-U_B]},\nonumber
\end{eqnarray}
where $\mathbf{a}$, $\mathbf{b}$, $\mathbf{c}$, $\mathbf{d}$ are column vectors with $\mathbf{a}_i=\langle A_i\rangle_k$, $\mathbf{b}_i=\langle B_i\rangle_k$, $\mathbf{c}_i=\langle A_i\rangle_k-\langle A_i\rangle_{k'}$, $\mathbf{d}_i=\langle B_i\rangle_k-\langle B_i\rangle_{k'}$, and we have used the convex property of the trace norm and the Cauchy-Schwarz inequality. It is worth noticing that the inequality (\ref{C}) can also be proved from the symmetric CMC using arbitrary local observables \cite{otfried,TLUR}, which can be seen from the proof of Proposition 1 in Ref. \cite{TLUR}. For the completeness of this work, we have here given direct proofs of the inequalities (\ref{R}) and (\ref{C}).

First, one practical entanglement condition can be derived from the inequality (\ref{C}).

\textit{Proposition 1.} (a) Consider the bipartite CV systems with $M$ modes in subsystem A and $N$ modes in subsystem B, respectively. One can choose the local operators $\{A_i\}_{i=1}^{2M}$ and $\{B_i\}_{i=1}^{2N}$ as follows,
\begin{align}
&\{A_i\}=\{a_1\hat{x}_1,a_2\hat{p}_1,\cdots,a_{2M-1}\hat{x}_M,a_{2M}\hat{p}_M\},\nonumber\\
&\{B_i\}=\{b_1\hat{x}_{M+1},b_2\hat{p}_{M+1},\cdots,b_{2N-1}\hat{x}_{M+N},b_{2N}\hat{p}_{M+N}\},\nonumber
\end{align}
where $a_i$, $b_i$ are real parameters. Therefore, every bipartite separable state must satisfy the inequality (\ref{C}) with
\begin{eqnarray}
&&U_A=|a_1a_2|+|a_3a_4|+\cdots+|a_{2M-1}a_{2M}|,\nonumber\\
&&U_B=|b_1b_2|+|b_3b_4|+\cdots+|b_{2N-1}b_{2N}|.\nonumber
\end{eqnarray}
(b) Specially, for two-mode CV systems, the following inequality must be satisfied by separable states,
\begin{eqnarray}
&&[\langle(\Delta\hat{x}_{1})^{2}\rangle+\langle(\Delta\hat{p}_{1})^{2}\rangle-1][\langle(\Delta\hat{x}_{2})^{2}\rangle+\langle(\Delta\hat{p}_{2})^{2}\rangle-1]\nonumber\\
&\geq&[(\langle\hat{x}_{1}\hat{x}_{2}\rangle-\langle\hat{x}_{1}\rangle\langle\hat{x}_{2}\rangle)\mp(\langle\hat{p}_{1}\hat{p}_{2}\rangle-\langle\hat{p}_{1}\rangle\langle\hat{p}_{2}\rangle)]^{2}\nonumber\\
&&+[(\langle\hat{x}_{1}\hat{p}_{2}\rangle-\langle\hat{x}_{1}\rangle\langle\hat{p}_{2}\rangle)\pm(\langle\hat{p}_{1}\hat{x}_{2}\rangle-\langle\hat{p}_{1}\rangle\langle\hat{x}_{2}\rangle)]^{2}.\label{pro1b}
\end{eqnarray}

\textit{Proof.} (a) According to the uncertainty relations $\langle(\Delta\hat{x}_{j})^{2}\rangle+\langle(\Delta\hat{p}_{j})^{2}\rangle\geq2\sqrt{\langle(\Delta\hat{x}_{j})^{2}\rangle\langle(\Delta\hat{p}_{j})^{2}\rangle}\geq|\langle[\hat{x}_{j},\hat{p}_{j}]\rangle|=1$,
one can obtain that $U_A=|a_1a_2|+|a_3a_4|+\cdots+|a_{2M-1}a_{2M}|$ and $U_B=|b_1b_2|+|b_3b_4|+\cdots+|b_{2N-1}b_{2N}|$, and every separable state must satisfy the inequality (\ref{C}). (b) Let us choose $\hat{A}_{1}=|a|\hat{x}_{1}$, $\hat{A}_{2}=|a|\hat{p}_{1}$, $\hat{B}_{1}=\hat{x}_{2}/a$, and $\hat{B}_{2}=-\hat{p}_{2}/a$, where $a$ is a nonzero real parameter. It follows that
$[\sum_i\langle(\Delta A_i)^2\rangle_{\rho_A}-U_A][\sum_j\langle(\Delta B_j)^2\rangle_{\rho_B}-U_B]=[\langle(\Delta\hat{x}_{1})^{2}\rangle+\langle(\Delta\hat{p}_{1})^{2}\rangle-1][\langle(\Delta\hat{x}_{2})^{2}\rangle+\langle(\Delta\hat{p}_{2})^{2}\rangle-1]$, and $\|C\|^{2}=[(\langle\hat{x}_{1}\hat{x}_{2}\rangle-\langle\hat{x}_{1}\rangle\langle\hat{x}_{2}\rangle)\mp(\langle\hat{p}_{1}\hat{p}_{2}\rangle-\langle\hat{p}_{1}\rangle\langle\hat{p}_{2}\rangle)]^{2}
+[(\langle\hat{x}_{1}\hat{p}_{2}\rangle-\langle\hat{x}_{1}\rangle\langle\hat{p}_{2}\rangle)\pm(\langle\hat{p}_{1}\hat{x}_{2}\rangle-\langle\hat{p}_{1}\rangle\langle\hat{x}_{2}\rangle)]^{2}$.
Therefore, the inequality (\ref{pro1b}) can be derived from the inequality (\ref{C}). \hfill $\square$

\textit{Remark.} It is worth noticing that the condition in Proposition 1(b) is stronger than the inequalities (\ref{Duan}) and (\ref{TLUR}). Moreover, Proposition 1(a) can be used to detect bound entangled Gaussian state.

\textit{Corollary 1.} Proposition 1(b) is stronger than the inequalities (\ref{Duan}) and (\ref{TLUR}).

\textit{Proof.} Based on the inequality (\ref{C}) and $[\sum_k(\langle\hat{A}_k\otimes\hat{B}_k\rangle-\langle\hat{A}_k\otimes I\rangle\langle I\otimes\hat{B}_k\rangle)]^2=(\mathrm{Tr}C)^2\leq\|C\|^2\leq[\sum_i\langle(\Delta \hat{A}_i)^2\rangle_{\rho_A}-U_A][\sum_j\langle(\Delta \hat{B}_j)^2\rangle_{\rho_B}-U_B]$, one can obtain that the inequality (\ref{C}) is strictly stronger than $\sqrt{[\sum_i\langle(\Delta \hat{A}_i)^2\rangle_{\rho_A}-U_A][\sum_j\langle(\Delta \hat{B}_j)^2\rangle_{\rho_B}-U_B]}\pm\sum_k(\langle\hat{A}_k\otimes\hat{B}_k\rangle-\langle\hat{A}_k\otimes I\rangle\langle I\otimes\hat{B}_k\rangle)\geq0$ which is actually Lemma 1 in Ref. \cite{TLUR}. Notice that Theorem 1 in Ref. \cite{TLUR} can be obtained from Lemma 1. Thus, the inequality (\ref{C}) is strictly stronger than Theorem 1 in Ref. \cite{TLUR}. Using the same local operators, $\hat{A}_{1}=|a|\hat{x}_{1}$, $\hat{A}_{2}=|a|\hat{p}_{1}$, $\hat{B}_{1}=\hat{x}_{2}/a$, and $\hat{B}_{2}=-\hat{p}_{2}/a$, the inequality (\ref{C}) can be written as Proposition 1(b), and Theorem 1 in Ref. \cite{TLUR} can be written as the inequality (\ref{TLUR}). Therefore, Proposition 1(b) turns out to be stronger than the inequality (\ref{TLUR}).
Notice that the inequality (\ref{TLUR}) is an improved form of the inequality (\ref{Duan}), hence, Corollary 1 holds. \hfill $\square$

\textit{Example 1.} Let us consider the bound entangled Gaussian state shown in Ref. \cite{Wolf}. The covariance matrix of this $2\times2$ bound entangled Gaussian state is
\begin{equation}\label{excov}
\gamma=\left(\begin{array}{cccccccc}
  2& 0& 0& 0& 1& 0& 0& 0\\
  0& 1& 0& 0& 0& 0& 0& -1\\
  0& 0& 2& 0& 0& 0& -1& 0\\
  0& 0& 0& 1& 0& -1& 0& 0\\
  1& 0& 0& 0& 2& 0& 0& 0\\
  0& 0& 0& -1& 0& 4& 0& 0\\
  0& 0& -1& 0& 0& 0& 2& 0\\
  0& -1& 0& 0& 0& 0& 0& 4
  \end{array}\right)\;.
\end{equation}
According to Proposition 1(a), one can choose $A_{1}=a_{1}\hat{x}_{1}$, $A_{2}=a_{2}\hat{p}_{1}$, $A_{3}=a_{3}\hat{x}_{2}$, $A_{4}=a_{4}\hat{p}_{2}$,  and $B_{1}=b_{1}\hat{x}_{3}$, $B_{2}=b_{2}\hat{p}_{3}$, $B_{3}=b_{3}\hat{x}_{4}$, $B_{4}=b_{4}\hat{p}_{4}$.
Using the covariance matrix in Eq.~(\ref{excov}), it follows that $\|C\|^{2}=(|a_{1}b_{1}|+|a_{2}b_{4}|+|a_{3}b_{3}|+|a_{4}b_{2}|)^{2}/4$, $\sum_i\langle(\Delta A_i)^2\rangle_{\rho_A}-U_A=a_{1}^{2}+a_{2}^{2}/2-|a_{1}a_{2}|+a_{3}^{2}+a_{4}^{2}/2-|a_{3}a_{4}|$, and $\sum_j\langle(\Delta B_j)^2\rangle_{\rho_B}-U_B=b_{1}^{2}+2b_{2}^{2}-|b_{1}b_{2}|+b_{3}^{2}+2b_{4}^{2}-|b_{3}b_{4}|$, where $U_A=|a_1 a_2|+|a_3 a_4|$ and $U_B=|b_1 b_2|+|b_3 b_4|$ are used.
If we choose $a_{1}=a_{3}=b_{2}=b_{4}=\sqrt{2}/2$ and $a_{2}=a_{4}=b_{1}=b_{3}=1$, we obtain
\begin{eqnarray}
&&\|C\|^{2}-[\sum_i\langle(\Delta A_i)^2\rangle_{\rho_A}-U_A][\sum_j\langle(\Delta B_j)^2\rangle_{\rho_B}-U_B]\nonumber\\
&=&6\sqrt{2}-8>0,\label{violate}
\end{eqnarray}
where the inequality (\ref{C}) via Proposition 1(a) is violated. Therefore, our criterion detects $\gamma$ as an entangled Gaussian state. Note that Ref. \cite{Wolf} has proved that the PPT criterion cannot detect its entanglement, which confirms that Proposition 1(a) is not a corollary of the PPT criterion.

The second practical condition can be derived from the inequality (\ref{R}), and it can be used to detect bound entangled non-Gaussian states.

\textit{Proposition 2.} For two-mode CV systems, one can choose the local operators as $I_N$ and generators of SU(N) $\{\lambda_i\}$, that is,
$\{A_i\}_{i=1}^{N^2}=\{I_{N}^{A}/\sqrt{N},\lambda_{1}^{A}/\sqrt{2},\cdots,\lambda_{N^2-1}^{A}/\sqrt{2}\}$ and $\{B_j\}_{j=1}^{N^2}=\{I_{N}^{B}/\sqrt{N},\lambda_{1}^{B}/\sqrt{2},\cdots,\lambda_{N^2-1}^{B}/\sqrt{2}\}$, where
$I_N$ is the projection to the Hilbert space of the truncated $N$ levels. Based on the inequality (\ref{R}), every separable state $\rho$ must satisfy the following inequality
\begin{equation}\label{pro2}
    \|R\|\leq\langle I_{N}^{A}\otimes I_{N}^{B}\rangle_{\rho}.
\end{equation}

\textit{Proof.} If one chooses the local operators $\{A_i\}$ and $\{B_j\}$ as $\{A_i\}_{i=1}^{N^2}=\{I_{N}^{A}/\sqrt{N},\lambda_{1}^{A}/\sqrt{2},\cdots,\lambda_{N^2-1}^{A}/\sqrt{2}\}$ and $\{B_j\}_{j=1}^{N^2}=\{I_{N}^{B}/\sqrt{N},\lambda_{1}^{B}/\sqrt{2},\cdots,\lambda_{N^2-1}^{B}/\sqrt{2}\}$, one obtains that $\sum_i\langle A_i\rangle_{k}^{2}=\mathrm{Tr}(I_{N}^{A} \rho_{k}^{A} I_{N}^{A})^2=(\mathrm{Tr}I_{N}^{A} \rho_{k}^{A})^2$ and $\sum_j\langle B_j\rangle_{k}^{2}=\mathrm{Tr}(I_{N}^{B} \rho_{k}^{B} I_{N}^{B})^2=(\mathrm{Tr}I_{N}^{B} \rho_{k}^{B})^2$, where the fact that $\rho_{k}^{A}$ and $\rho_{k}^{B}$ are pure states is used. According to the inequality (\ref{R}), it follows that $\|R\|\leq\sum_{k}p_k\sqrt{(\sum_i\langle A_i\rangle_{k}^{2})(\sum_j\langle B_j\rangle_{k}^{2})}=\sum_k p_k (\mathrm{Tr}I_{N}^{A} \rho_{k}^{A})(\mathrm{Tr}I_{N}^{B} \rho_{k}^{B})=\langle I_{N}^{A}\otimes I_{N}^{B}\rangle_{\rho}$.  \hfill $\square$

\textit{Remark.} Proposition 2 actually shows that Eq. (\ref{R}) still holds for separable states in the Hilbert space of the truncated $N$ levels, and it coincides with the fact that if a state $\rho$ by means of locally projecting onto a subspace results an entangled state $\rho'$, the state $\rho$ is also entangled. The advantage of Proposition 2 is that Eq. (\ref{pro2}) can be experimentally measured in principle.

\begin{figure}
\begin{center}
\includegraphics[scale=0.8]{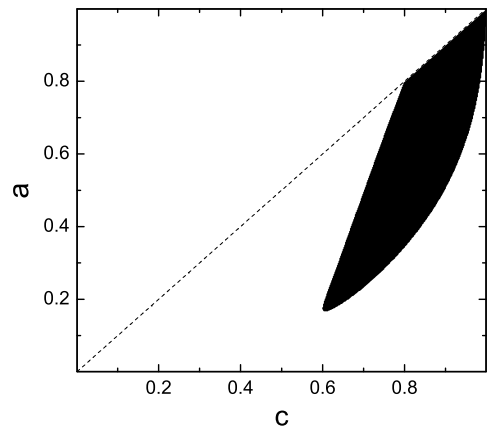}
\caption{Detection of the bound entangled non-Gaussian states in Eq. (\ref{nonGaussian}) with $a_n=a^n$, $c_n=c^n$ for the range $0<a<c<1$. These states occupy the region below the dotted line and the states in the black-shaded region can be detected as entangled by Proposition 2.}\label{1}
\end{center}
\end{figure}

\textit{Example 2.} Let us consider the bound entangled non-Gaussian state shown in Ref. \cite{Horodecki2000}. Horodecki and Lewenstein have introduced a class of CV bound entangled states,
\begin{equation}\label{nonGaussian}
    \rho=\frac{1}{A}\bigg(|\Psi\rangle\langle\Psi|+\sum_{n=1}^{\infty}\sum_{m>n}^{\infty}|\Psi_{mn}\rangle\langle\Psi_{mn}|\bigg),
\end{equation}
where $|\Psi\rangle=\sum_{n=1}^{\infty}a_n|n,n\rangle$ and $|\Psi_{mn}\rangle=c_m a_n|n,m\rangle+(c_m)^{-1}a_m|m,n\rangle$, for $n<m$ with complex $a_n$ and $c_n$ such that $0<|c_{n+1}|<|c_n|<1$, and the normalizing factor $A=\|\Psi\|^2+\sum_{n=1}^{\infty}\sum_{m>n}^{\infty}\|\Psi_{mn}\|^2$. For simplicity, we consider the case of $a_n=a^n$, $c_n=c^n$ for $0<a<c<1$. Ref. \cite{Horodecki2000} shows that the above state satisfies $\rho=\rho^{T_B}$ and has the PPT property. Based on Proposition 2, we choose $I_3$ and eight Gell-Mann matrices $\{G_i\}_{i=1}^{8}$ as our local operators, i.e. $\{A_i\}=\{I_3^A/\sqrt{3},G_1^A/\sqrt{2},\cdots,G_8^A/\sqrt{2}\}$ and $\{B_j\}=\{I_3^B/\sqrt{3},G_1^B/\sqrt{2},\cdots,G_8^B/\sqrt{2}\}$, where $I_3=|1\rangle\langle1|+|2\rangle\langle2|+|3\rangle\langle3|$, $G_1=|1\rangle\langle2|+|2\rangle\langle1|$, $G_2=-i(|1\rangle\langle2|-|2\rangle\langle1|)$, $G_3=|1\rangle\langle1|-|2\rangle\langle2|$, $G_4=|1\rangle\langle3|+|3\rangle\langle1|$, $G_5=-i(|1\rangle\langle3|-|3\rangle\langle1|)$, $G_6=|2\rangle\langle3|+|3\rangle\langle2|$, $G_7=-i(|2\rangle\langle3|-|3\rangle\langle2|)$, $G_8=(|1\rangle\langle1|+|2\rangle\langle2|-2|3\rangle\langle3|)/\sqrt{3}$. One can see that $\|R\|/\langle I_{3}^{A}\otimes I_{3}^{B}\rangle_{\rho}=1.04761>1$ for $a=0.5$ and $c=0.8$, which violates the inequality (\ref{pro2}). Moreover, using the same local operators, we have checked all the states with $0<a<c<1$ by Proposition 2, and the results are shown in Fig. \ref{1}. Although not all the states with $0<a<c<1$ can be detected, the states in the black-shaded region in Fig. \ref{1} can be detected as entangled states by our condition. As mentioned above, all the states of Eq. (\ref{nonGaussian}) have the PPT property. Thus, we see that Proposition 2, which is not a corollary of the PPT criterion, can be used to detect bound entangled non-Gaussian states.

\section{Discussion and conclusion}
Notice that Proposition 1 can be generalized when one use other local operators instead of $\hat{x}_i$, $\hat{p}_i$ with real parameters. For instance, Refs. \cite{multi1,multi2} introduced multiphoton operators $\hat{A}^{(k)\dag}=\sqrt{[[\hat{N}/k]](\hat{N}-k)!/\hat{N}!}\hat{a}^{\dag k}$, satisfying $[\hat{A}^{(k)}, \hat{A}^{(k)\dag}]=\eins$, where $\hat{N}=\hat{a}^{\dag}\hat{a}$, $[[\hat{N}/k]]=\sum_{n}[[n/k]]|n\rangle\langle n|$, $(\hat{N}-k)!/\hat{N}!=\sum_{n}[(n-k)!/n!]|n\rangle\langle n|$, and $[[n/k]]$ denotes the largest positive integer less than or equal to $n/k$. One can define $\hat{X}^{(k)}_{j}=(\hat{A}^{(k)\dag}+\hat{A}^{(k)})/\sqrt{2}$ and $\hat{P}^{(k)}_{j}=i(\hat{A}^{(k)\dag}-\hat{A}^{(k)})/\sqrt{2}$, and use them in Proposition 1 instead of $\hat{x}_j$, $\hat{p}_j$.

There are still several questions about Propositions 1 and 2. First, we use real parameters $a_i$ and $b_i$ in Proposition 1(a), and simply choose $a_{1}=a_{3}=b_{2}=b_{4}=\sqrt{2}/2$ and $a_{2}=a_{4}=b_{1}=b_{3}=1$ in Example 1. For a given state, how can one choose real parameters $a_i$ and $b_i$ in order to optimize the effectiveness of Proposition 1(a)? This question is interesting and worth for further research. Second, it can be seen from Fig. \ref{1} that the states in the white region of the lower triangle cannot be detected by Proposition 2. One has to propose other CV entanglement conditions for these states. Third, the generators of SU(N) have been used in Proposition 2, which, compared with $\hat{x}$ and $\hat{p}$, are not easy to directly implement in experiment. Finally, there are numerous bound entangled states in discrete variable systems. However, only few examples of bound entangled states exist in CV variable systems, like Examples 1 and 2. Therefore, a new class of bound entangled states for CVs must be constructed for testing general entanglement conditions.

In conclusion, we have presented entanglement criteria in order to detect bound entangled states for CVs. Using these criteria, we have shown that bound entangled Gaussian state and bound entangled non-Gaussian state proposed in Refs. \cite{Wolf,Horodecki2000} can be detected.

\section*{ACKNOWLEDGMENTS}
We would like to thank O. G\"uhne and Sixia Yu for helpful discussions and gratefully acknowledge valuable comments by W. Vogel, J. Eisert and A. Miranowicz. This work was funded by the National Fundamental Research Program (Grant No. 2006CB921900), the National Natural Science Foundation of China (Grants No. 10674127, No. 60621064 and No. 10974192), the Innovation Funds from the Chinese Academy of Sciences, and the K.C. Wong Foundation. H.N. is supported by an NPRP grant 1-7-7-6 from Qatar National Research Funds. C.J.Z. acknowledges the financial support of ASTAR Grant R-144-000-189-305.

\end{document}